\documentclass[prl,amsmath,twocolumn,amssymb,aps,superscriptaddress,floatfix,preprintnumbers,10pt,notitlepage]{revtex4-1}
\usepackage{times}
\usepackage{graphicx}
\usepackage{amsmath, braket, amsfonts}
\usepackage{amssymb}
\usepackage{natbib}
\usepackage{sidecap}
\usepackage{bm, color, ulem}
\usepackage{xcolor} % HZ: temporarily added to enable more text colors that are comfortable to read

\def\lsim{\lower -0.3ex \hbox{$<$} \kern -0.75em \lower 0.7ex \hbox{$\sim$}}
\def\gsim{\lower -0.3ex \hbox{$>$} \kern -0.75em \lower 0.7ex \hbox{$\sim$}}

\begin{document}
\title{Skyrmion solids in monolayer graphene}
\author{H. Zhou}
\affiliation{Department of Physics, University of California, Santa Barbara CA 93106 USA}
\author{H. Polshyn}
\affiliation{Department of Physics, University of California, Santa Barbara CA 93106 USA}
\author{T. Taniguchi}   
\affiliation{National Institute for Materials Science, Tsukuba, Ibaraki 305-0044, Japan}
\author{K. Watanabe}
\affiliation{National Institute for Materials Science, Tsukuba, Ibaraki 305-0044, Japan}
\author{A.F. Young}
\affiliation{Department of Physics, University of California, Santa Barbara CA 93106 USA}
\date{\today}
 \maketitle

\textbf{Partially filled Landau levels host competing orders, with electron solids prevailing close to integer fillings before giving way to fractional quantum Hall liquids as the Landau level fills\cite{lam_liquid-solid_1984,levesque_crystallization_1984}.  Here, we report the observation of an electron solid with noncolinear spin texture in monolayer graphene, consistent with solidification of skyrmions\cite{brey_skyrme_1995}---topological spin textures characterized by quantized electrical charge\cite{sondhi_skyrmions_1993,fertig_charged_1994}. In electrical transport, electron solids reveal themselves through a rapid metal-insulator transition in the bulk electrical conductivity as the temperature is lowered\cite{jiang_quantum_1990}, accompanied by the emergence of strongly nonlinear dependence on the applied bias voltage\cite{goldman_evidence_1990,jiang_magnetotransport_1991}. We probe the spin texture of the solids using a modified Corbino geometry\cite{polshyn_quantitative_2018,zeng_high_2018} that allows ferromagnetic magnons to be launched and detected\cite{wei_electrical_2018,stepanov_long-distance_2018}.
We find that magnon transport is highly efficient when one Landau level is filled ($\nu=1$), consistent with quantum Hall ferromagnetic spin polarization; however, even minimal doping immediately quenches the the magnon signal while leaving the vanishing low-temperature charge conductivity unchanged.  Our results can be understood by the formation of a solid of charged skyrmions near $\nu=1$, whose noncolinear spin texture leads to rapid magnon decay. Data near fractional fillings further shows evidence for several fractional skyrmion crystals, suggesting that graphene hosts a vast and highly tunable landscape of coupled spin and charge orders.}

When a spin-degenerate Landau level is half filled (corresponding to filling factor $\nu=2\pi\ell_B^2 n=1$ 
with the magnetic length $\ell_B=(eB_\perp/\hbar)^{-1/2}$ and  $n$ the carrier density), dominant exchange interactions drive the system into an insulating quantum Hall ferromagnetic (QHFM) state\cite{girvin_quantum_1999}.
Notably, only minimal anisotropy is introduced by the Zeeman energy ($E_Z=g\mu_B B_T$), which is much smaller than the exchange energy ($E_X=\sqrt{\frac{\pi}{2}}\frac{e^2}{\epsilon \ell_B}$). As a result, smooth skyrmion spin textures, which minimize exchange energy at the expense of reversing more spins relative to the external magnetic field, can be energetically favorable\cite{sondhi_skyrmions_1993}.  In contrast to skyrmions in metallic magnets, quantum Hall skyrmions have quantized electrical charge, leading to strong inter-skyrmion interactions and predicted spatial ordering of finite density skyrmions\cite{brey_skyrme_1995}. While thermal measurements\cite{bayot_giant_1996, bayot_critical_1997, melinte_heat_1999},  microwave spectroscopy\cite{zhu_pinning-mode_2010}, nuclear magnetic resonance and polarized absorption spectroscopy\cite{barrett_optically_1995, aifer_evidence_1996, smet_gate-voltage_2002, desrat_resistively_2002, gervais_evidence_2005, tracy_resistively_2006,mitrovic_nmr_2007,tiemann_nmr_2014, piot_disorder-induced_2016} all show  signatures of skyrmion solids, many features of these states of matter remain unexplored, including their degree of crystalline order and the nature of their elementary excitations. 

Owing to its wide range of electrostatic tunability and experimental accessibility, graphene provides a potentially ideal platform for probing the phase diagram of skyrmion solids. For example, the exposed surface may allow direct magnetic imaging, and the high sample quality  permits the exploration of skyrmion ground states over a broad range of electron density, magnetic field, and single particle wave function structure.
However, conventional detection schemes that rely on coupling to nuclear spins are challenging in carbon $\pi$-orbitals, and evidence to date for skyrmion physics in graphene has been limited to their indirect impact on measured energy gaps
\cite{dean_multicomponent_2011,young_spin_2012,feldman_unconventional_2012}. Indeed, while electron solids were recently reported in graphene in higher Landau levels\cite{chen_competing_2019}, no evidence has been reported for electron solidification of any kind within the lowest Landau level where skyrmion solids are theoretically anticipated.

\begin{figure*}[t!]
    \centering
    \includegraphics[width=183mm]{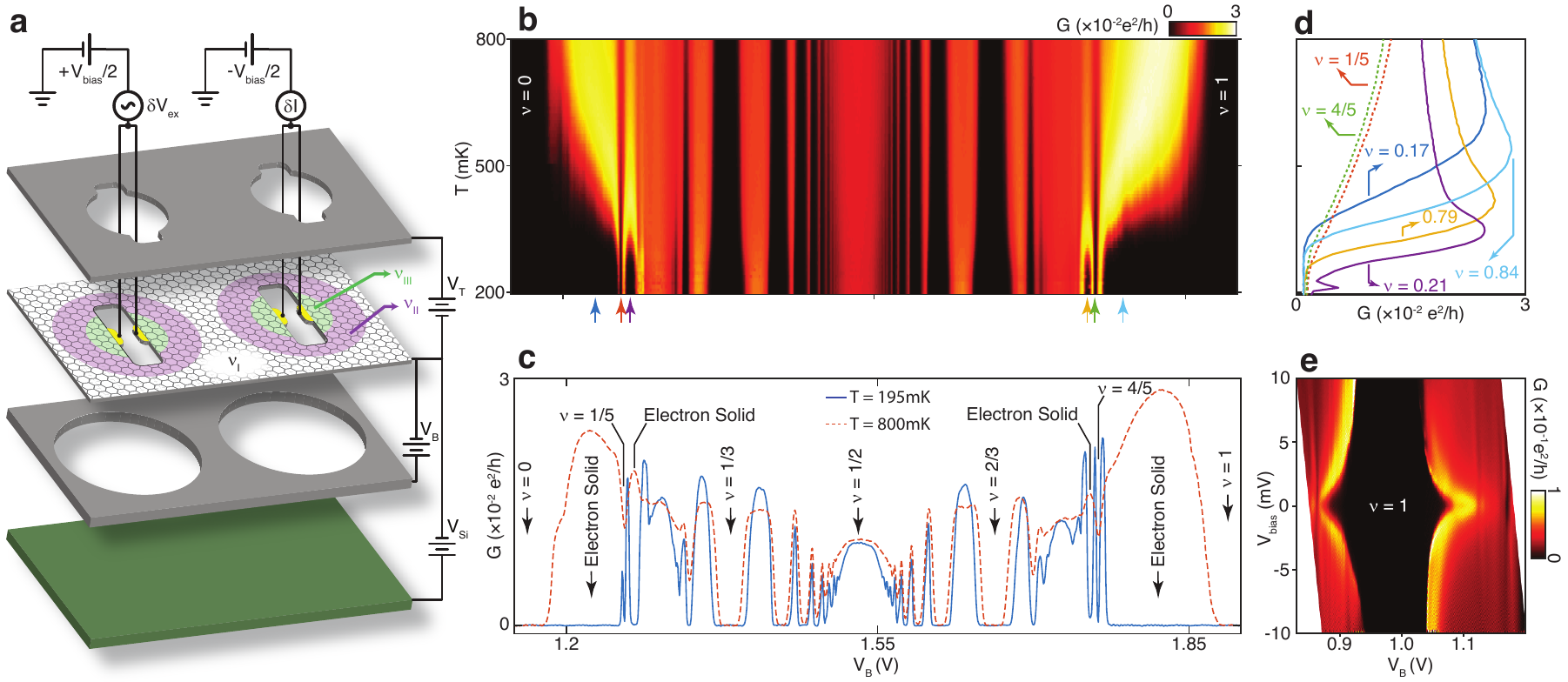}
     \caption{
     \textbf{Electron solid phases in the lowest Landau level.}
     % subfigure a
       % subfigure b
     \textbf{a}, Device and measurement schematic.  Voltages applied to graphite top- and bottom-gates ($V_\text{T}$ and $V_\text{B}$) and an additional silicon gate ($V_\text{Si}$) allow independent control of $\nu$ in regions I (white), II (pink), and III (green).  The differential conductance $G=\delta I/\delta V_\text{ex}$, proportional to the bulk conductivity of region I, is measured as shown, with  $\delta V_\text{ex}$ an applied low frequency voltage excitation and $\delta I$ the induced current across the $\nu_\text{I}$ region. The DC bias $V_\text{bias}=0$ unless indicated.
     % subfigure c
     \textbf{b}, Temperature-dependent $G$ for $0<\nu<1$ at $B_\perp = 13.5$T.
     % subfigure d
     \textbf{c}, $G$ measured at $T = 195$mK and $T = 800$mK. Labels highlight a subset of the observed insulating states, including FQH and electron solid phases. 
     % subfigure e
     \textbf{e}, $G$ measured as a function of $T$ for several values of $V_\text{B}$, indicated by colored arrows in panel \textbf{c}. 
     The hallmark of the electron solid phases is a sign reversal in $dG/dT$. 
     % subfigure f
     \textbf{f}, 
     $G$ near $\nu = 1$ at $B_\perp = 10$T measured as a function of $V_\text{B}$ and $V_\text{bias}$.  
     } % end of \caption
    \label{fig:fig1}
\end{figure*}

To explore the phase diagram of electron solids in graphene, we study ultra-high quality monolayer graphene devices fabricated in a Corbino geometry (see Methods). 
Two patterned graphite local gates and a doped p-Si global gate provide independent control of carrier density in two contact regions (II and III, rendered in pink and green in Fig. \ref{fig:fig1}a), in addition to the bottom gated bulk of region I. The electrical conductance, $G$, between two contact `islands' is directly proportional to the bulk electrical conductivity $\sigma_{xx}$.
Figs. \ref{fig:fig1}b-d show temperature dependent measurements of $G$ for $0<\nu<1$. Deep sequences of fractional quantum Hall(FQH) states manifest as conductance minima at fractional $\nu=\frac{p}{mp\pm1}$ and $\nu=1-\frac{p}{mp\pm1}$ with $m=2,4$, corresponding to the two-flux and four-flux composite fermion sequences. 
The data also feature anomalous insulating states near $\nu=0$ and $\nu=1$, extending beyond the $\nu=1/5$ and $4/5$ FQH states, respectively, which interrupt this insulating behavior. In contrast to even weak FQH states, which show a monotonic, simply activated temperature dependence up to temperatures of several Kelvin, the conductivity of the anomalous insulators transitions from metallic ($dG/dT<0$) to insulating ($dG/dT>0$) temperature coefficient at $T\sim300-600$~mK (Fig. \ref{fig:fig1}d).

% Fig. 1

% Fig. 2
\begin{figure*}[t!]
    \centering
    \includegraphics[width=183mm]{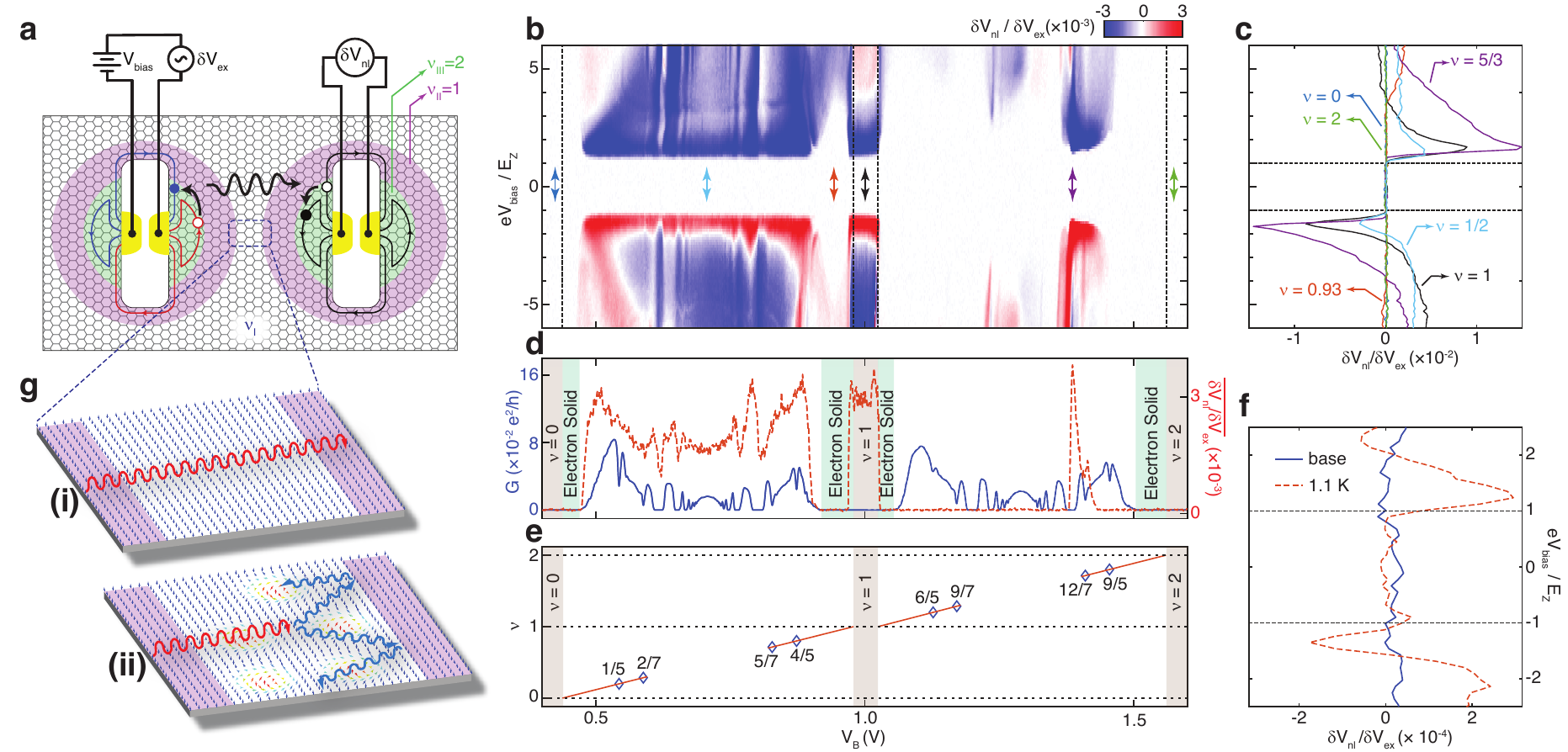}
    \caption{\textbf{Skyrmion solid phase near $\nu = 1$.}
    % subfigure a
    \textbf{a}, Schematic for measuring magnon transport.  Nonlocal measurements are performed with $\nu_\text{III}=2$ and $\nu_\text{II}=1$.  Applying $eV_\text{bias}>E_Z$ launches magnons through region II, in which magnons can propagate freely. The charge insulating bulk of region II prevents electron flow for all $\nu_\text{I}$.  Magnons transmitted through region I and back into the detector are absorved generating the measured nonlocal voltage.   
    % subfigure b
    \textbf{b}, False-color plot of the non-local response for $0<\nu_\text{I}<2$ as function of $eV_\text{bias}/E_\text{Z}$ and $V_\text{B}$ at $B_\perp = 10{\rm T}$.
    % subfigure c
    \textbf{c}, Non-local response as a function of $eV_\text{bias}/E_Z$ for different values of $V_\text{B}$, corresponding to the colored arrows in panel \textbf{a}. The dashed lines mark the Zeeman threshold $eV_\text{bias}/E_Z = \pm 1$, and approximate $\nu$ for each curve are indicated. 
    % subfigure d
    \textbf{d}, Blue: Local conductance $G$ for $0<\nu<2$. Red: Root-mean-square average of the non-local response in the range $-1.5E_Z<eV_\text{bias}<1.5E_Z$.  Boundaries between integer quantum Hall gaps and electron solids are indicated by the shaded regions.
    % subfigure e
    \textbf{e}, The symbols denote $\nu$ as determined from the positions in $V_\text{B}$ of FQH states associated with $\nu_\text{I}=1/5$, $2/7$, $5/7$, $4/5$, $6/5$, $9/7$, $12/7$ and $9/5$ as function of $V_\text{B}$. Red lines represent extrapolations of $\nu$ to the nearest integer determined using pairs of adjacent FQH states.  Shaded regions denote the finite extent in $V_\text{B}$ of states with integer $\nu$.  % subfigure f
 Nonlocal response at base temperature (blue) and T=1.1K (red) near $\nu=1$.  Lines represent data averages over a range of $0.927<V_\text{B} <0.955$V. 
    % subfigure g
    \textbf{g}, Magnon decay in a skyrmion solid. (i) When $\nu_I=1$, QHFM magnons are transmitted with little loss. (ii) When region I is in a skyrmion solid phase, magnons rapidly decay into other magnetic modes with $E<E_Z$ and spin less than one.  These modes cannot enter the  $\nu_\text{II}=1$ detector region, leading to suppression of the nonlocal response.
    } % end of \caption
    \label{fig:fig2}
\end{figure*}

The temperature dependence and density range of the anomalous insulators strongly resemble results obtained in clean semiconductor quantum wells\cite{jiang_quantum_1990,liu_observation_2012}, attributed to charge solidification at low quasiparticle density.  
In semiconductors, solidification was shown to be accompanied by the onset of nonlinear transport at anomalously low threshold electric fields, interpreted as depinning of the solid phases\cite{goldman_evidence_1990,jiang_magnetotransport_1991}.  
We observe similar behavior in a range of fillings near $\nu=1$ (Fig.~\ref{fig:fig1}e), and near other integer $\nu$ in the lowest Landau level (Fig.~\ref{fig:supp:nonlinear}).  Whereas the insulating states at integer and fractional fillings are highly robust to applied bias, the anomalous insulators near integer $\nu$ rapidly vanish at low bias voltages $V_\text{bias}\sim 1$mV. Taken together, the temperature and bias dependent transport data show that electron solid phases are a generic feature of the graphene lowest Landau level.  

While the electron solids proximal to $\nu=\pm2$, $\nu=\pm1$, and $\nu=0$ show only quantitative differences in their electrical transport properties, they are predicted to arise from solidification of quasiparticles with qualitatively different spin textures. In the case of $\nu=\pm2$, corresponding to a completely empty or full Landau level, one expects solidification of bare electrons or holes.  While the electron solid is thought to have a ferromagnetic ground state\cite{cote_spin-ordering_1996}, exchange interactions are weaker than typical experimental temperatures, making the physically realized state paramagnetic. Near charge neutrality, in contrast, the spin texture of the electron solid should depend on the nature of the $\nu=0$ ground state, which is thought to be a canted antiferromagnet or charge density wave in graphene samples with sublattice symmetry breaking\cite{kharitonov_phase_2012} (see Fig. \ref{fig:supp:transport}).  In the former case, noncolinear spin textures are expected for some range of $\nu$, while the latter should be paramagnetic by analogy with $\nu=\pm2$.  Finally, at $\nu=\pm1$ the low energy charged excitations are spin reversals, whose nature depends on the ratio $\kappa=E_Z/E_X$.  At large $\kappa$, the electron solid polarizes in a single spin branch; however, for sufficiently low $\kappa$, the spin-reversed excitations are skyrmions, and the solid should have long wavelength noncolinear spin texture\cite{brey_skyrme_1995}.  

To probe the spin texture of the electron solids, we measure the bulk transport of magnons\cite{wei_electrical_2018,stepanov_long-distance_2018}, as depicted schematically in Fig.~\ref{fig:fig2}a.  
Magnon transport is measured by setting $\nu_\text{III}=2$ and $\nu_\text{II}=1$, creating lateral heterojunctions near each pair of contacts, which serve as a magnon launcher and magnon detector, respectively. Applying a voltage bias $eV_\text{bias}\geq E_Z$ across contacts in the injector island launches neutral magnons into the electrically insulating $\nu_\text{II}=1$ region.
The probability of transporting a magnon across region I is then encoded in the nonlocal response measured across the $\nu_\text{I}$ sample bulk, $\delta V_\text{nl}/\delta V_\text{ex}$, where $\delta V_\text{ex}$ is a small AC voltage applied in series with $V_\text{bias}$ across the injector and $\delta V_\text{nl}$ is the AC voltage measured across the detector.
Note that in this measurement, the relevant magnons are excitations of the $\nu_\text{II}=1$ QHFM, which have quantized spin of one in the direction anti-parallel to the applied magnetic field and energy dispersion $E(k)=E_Z+\alpha k^2$\cite{kallin_excitations_1984,sondhi_skyrmions_1993,wei_electrical_2018}.  The nonlocal response
for $eV_\text{bias}$ slightly larger than $E_Z$ thus measures the probability of transmitting low-k magnons from injector to detector through region I. This signal can be suppressed either by the absence of compatible neutral modes in the intervening region, or by the presence of additional decay channels by which incoming magnons can relax their energy and momentum outside the spin system.

Figs. \ref{fig:fig2}b-c show data taken in this configuration. For all $\nu_\text{I}$, $\delta V_\text{nl}=0$ for $e V_\text{bias}<E_Z$, indicating that the measured response indeed arises from magnon transport.  No nonlocal responses are observed in the nonmagnetic $\nu_\text{I}=2$ state or in the $\nu=0$ state. Among the strongest nonlocal responses are observed at $\nu_\text{I}=1$ and $5/3$.  This is consistent with theoretical expectations of a fully spin-polarized QHFM (for $\nu=1$) and fully spin-polarized fractional QHFM (at $\nu=5/3$), both of which are predicted to admit long-distance transport of ferromagnetic magnons\cite{macdonald_magnons_1998}. 
The nonlocal response is not correlated with the charge conductivity (Fig. \ref{fig:fig2}d).  While the two regions $0<\nu<1$ and $1<\nu<2$ show similar sequences of FQH states, they differ markedly in their nonlocal response, reflecting differing ground state spin structure across the lowest Landau level.

% Fig. 3
\begin{figure}[htbp]
    \centering
    \includegraphics[width=89mm]{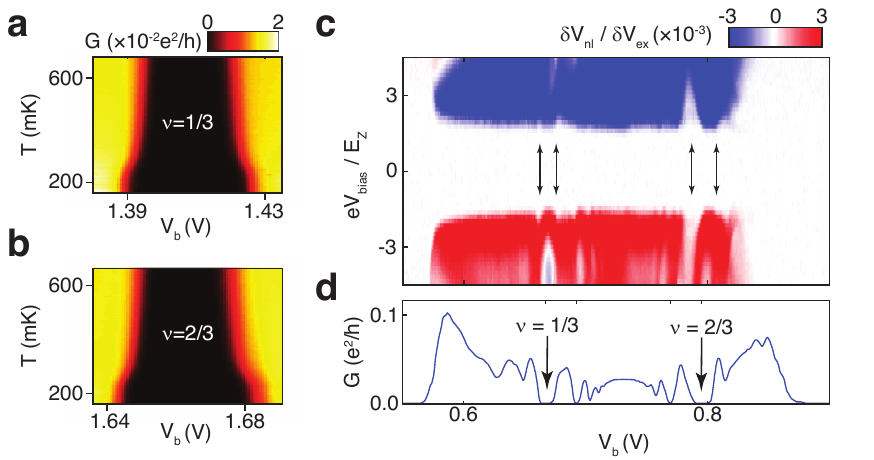}
\caption{\textbf{Evidence for fractional Skyrmion solid phases.}
% subfigure a and b
\textbf{a} Temperature-dependent conductance near $\nu = 1/3$  and (\textbf{b}) $\nu = 2/3$, showing a low-temperature transition to an insulator at small values of $|\nu-1/3|$ and $|\nu-2/3|$. 
% subfigure c
\textbf{c}, Non-local voltage for $0<\nu<1$ as a function of $V_\text{B}$ and $V_\text{bias}$ at $B_\perp = 7\rm T$. Arrows point indicate regions of suppressed  non-local response near the Zeeman threshold, reminiscent of the skyrmion signature observed near $\nu=1$. 
% subfigure d
\textbf{d}, Conductance within the same carrier density range as (\textbf{a}).
}
\label{fig:fig3}
\end{figure}

To determine the boundaries of the electron solid in Fig. \ref{fig:fig2}b, we calibrate the electron density using weakly formed FQH states from the four flux sequence (Fig. \ref{fig:fig2}e).  This analysis shows an electron solid regime of vanishing low-temperature $G$ but finite quasiparticle density near each integer $\nu$, consistent with the temperature dependent and nonlinear transport measurements described above.  
Comparison with the nonlocal response shows that all four electron solids suppress magnon transport. Near $\nu=2$, this is consistent with a paramagnetic electron solid, while at $\nu=0$ the solid presumably has noncollinear spin textures or is also a paramagnet.  Notably, in both cases the electron solid is not distinguishable from its parent integer state by either its low temperature charge conductivity or its magnon transport, both of which vanish. This is in sharp contrast to the solids surrounding $\nu=1$, which do not allow magnon transport in contrast to their quantum Hall ferromagnetic parent state.  
Remarkably, the boundary of magnon transport aligns precisely with the edges of the $\nu=1$ gap determined from quantum capacitance considerations (see also Fig. \ref{fig:supp:nu1gap}). 
We thus conclude that while the QHFM at exactly $\nu_\text{I}=1$ conducts magnons, a distinct insulating state characterized by the total suppression of magnon transport obtains for $0.89\lesssim \nu_\text{I}<1$ on the hole side and $1<\nu_\text{I}\lesssim 1.06$ on the electron side of $\nu=1$.  
To within experimental error, arbitrarily small doping of the $\nu_\text{I}=1$ state appears sufficient to fully suppress magnon transport. At high temperatures, where charge conductivity indicates the electron solid has melted, magnon transport is restored (Fig. \ref{fig:fig2}f).  

We propose that the suppression of magnon transport in the electron solid near $\nu_\text{I}=1$ implies that charge solidification is accompanied by noncolinear spin textures---the low-doping insulating phase is a skyrmion solid, rather than a spin polarized electron solid. In the skyrmion solid phase, dilute charge carriers are localized by a combination of their mutual Coulomb repulsion and, potentially, disorder, while the electron spins are driven by exchange interactions into large-scale noncolinear ground state spin textures.
Fig. \ref{fig:fig2}g outlines a mechanism for the resulting suppression of magnon transport.  Magnons are launched through the $\nu_\text{II}=1$ region, where they are gapped ($E_\text{M}(k)\geq E_\text{Z}$, with $E_\text{M}(k)$ the magnon dispersion) 
and characterized by quantized spin. However, within the skyrmion solid phase, the excitation spectrum includes spin modes with $E<E_Z$\cite{cote_collective_1997}. When a magnon with energy $E_Z$ enters the skyrmion solid phase, it can decay to one or more gapless spin modes (as well as one or more spinless phonons).  Because a magnon with $E<E_Z$ cannot enter the $\nu=1$ region surrounding the detector, the spins will be trapped in the skyrmion solid phase. The low energy magnons ultimately relax their energy and angular momentum outside the spin system, for example at the sample boundary, and generate no nonlocal voltage in the detector region.

Electron solids are also predicted to occur in the vicinity of FQH states\cite{archer_static_2011}, and evidence for such phases was recently reported in microwave spectroscopy experiments\cite{zhu_observation_2010}. Figs. \ref{fig:fig3}a and b show the low temperature behavior of $G$ at $B_{\perp}=13.5$T in the vicinity of  $\nu=1/3$ and $2/3$.  As is the case near integer $\nu$, $G$ undergoes a metal-insulator transition at $T\approx$300mK, suggestive of new solid phases. Intriguingly, the strong nonlocal response at precise fractional filling is subtly suppressed upon doping (\ref{fig:fig3}c and d).  Subtle features associated with an increased threshold for nonlocal response appear adjacent to the FQH states. This is reminiscent of the response of the skyrmion solids near $\nu=1$, albeit with a dramatically lower threshold $V_\text{bias}$ for the restoration of nonlocal response. Our data thus suggest that noncollinear spin-textured ground states of fractional skyrmions may prevail near these fillings.
We expect that additional experiments at lower temperatures and magnetic field (see Fig. \ref{fig:supp:heating}) and in cleaner samples may reveal a plethora of additional spin textured phases, allowing the interacting phase diagram of two dimensional electrons to be mapped in great detail.

\section{Methods}

Devices were fabricated using a dry transfer procedure, following Refs. \cite{zeng_high_2018,polshyn_quantitative_2018}. An optical image of the device discussed in the main text is shown in Fig. \ref{fig:supp:optical_image}.  The monolayer graphene (MLG), graphite, and hexagonal-boron nitride (hBN) flakes were prepared by mechanical exfoliation. In the device described in the main text, the thicknesses of the hBN flakes above and below the MLG flake are 34 nm and 40 nm respectively.
Top and bottom graphite gates were patterned as shown in Fig. \ref{fig:fig2}a using plasma etching. 
The chosen shape of the gates yields three distinct MLG  regions; the carrier density in each of them can be controlled independently.
Region I (rendered in white in Fig. \ref{fig:fig1}a) is exposed to and therefore controlled by the top and bottom graphite gates; region II (rendered in pink) is controlled by the top graphite gate and silicon gate; region III (rendered in green) is controlled by the silicon gate alone. 
In contrast to prior work, in our devices each internal `island,' defined as a region completely surrounded by sample bulk at filling $\nu_I$, features two separate contacts which enable both calibration of the filling factors $\nu_{II}$ and $\nu_{III}$ as well as the injection and  detection of magnons. 

Transport measurements of the device were done in a dilution refrigerator equipped with a 14T superconducting magnet and heavy RF and audio frequency filtering with a cutoff frequency of $\sim$ 10 kHz. The measurements were performed at base temperature unless indicated, corresponding to a measured temperature of $\lesssim$ 20 mK on the probe. The temperature-dependent  measurements were done by controlling the temperature  using a heater mounted on a mixing chamber plate.

Fig.~\ref{fig:supp:wiring}a shows a schematic measurement of the conductance $G$, proportional to the conductivity in region I.  While the device design ensures edge state coupling between the contact and the boundary of region II, considerable care must be taken to ensure that the measured $G$ is not dominated by spurious series resistances arising from the I/II interface.   To address this source of systematic error, the filling factor of region II is tuned to a bulk conducting state using $V_\text{T}$.  
The value of $V_\text{T}$ that maximizes the conductivity of region II is found from a two-dimensional sweep of $V_\text{T}$ and $V_\text{B}$ with silicon gate voltage $V_\text{Si}$ fixed. In practice, we find that setting $|\nu_\text{I} - \nu_\text{II}|<1$ avoids high series resistances from an insulating I/II interface.  
$\nu_I$ is then controlled using the bottom gate voltage $V_\text{B}$.  Fig.~\ref{fig:supp:wiring}b shows a typical trace of $G$ as a function of $V_\text{B}$ at $B_{\perp} = 8$T.  
The differential conductance was measured using a lock-in amplifier with a 100 $\mu$V excitation at 17.777 Hz.

The measurements of the non-local response due to magnon transport require precise tuning of regions II and III to $\nu_\text{II} = 1$ and $\nu_\text{III} = 2$.  To find the required values of the gate voltages $V^*_\text{T}$ and $V^*_\text{Si}$ for a given $B_{\perp}$ we first measure the intra-island conductance, as shown in Fig. \ref{fig:supp:wiring}c, as a function of $V_\text{Si}$ with $V_\text{T}$ and $V_\text{B}$ set to zero. We choose a value of $V^*_\text{Si}$ at which $G = 2e^2/h$, ensuring $\nu_{III}\approx 2$. Next, we measure intra-island conductance as a function of $V_\text{T}$ at $V_\text{Si}=V^*_\text{Si}$, $V_\text{B}=0$ and pick the value $V^*_\text{T}$ at which $G = e^2/h$ (Fig. \ref{fig:supp:wiring}d).  $V_T^*$ and $V_{Si}^*$ are then fixed for the nonlocal measurements described in the main text.  Nonlocal measurements are performed with $V_\text{ex} \approx 0.075V_Z$ at 1234.5 Hz. The higher frequency significantly reduced the noise level while introducing negligible phase shift.

\section{Acknowledgments: }
The authors acknowledge discussions with B. Halperin, C. Huang, A. Macdonald, and M. Zalatel.
Experimental work at UCSB was supported by the Army Research Office under awards MURI W911NF-16-1-0361 and W911NF-16-1-0482. KW and TT acknowledge support from the Elemental Strategy Initiative conducted by the MEXT, Japan and and the CREST (JPMJCR15F3), JST.
AFY acknowledges the support of the David and Lucile Packard Foundation and and Alfred. P. Sloan Foundation.

\section{Author Contributions: }
HZ and HP fabricated the devices.  HZ, HP and AFY performed the measurements and analyzed the data. HZ and AFY wrote the manuscript.  TT and KW grew the hexagonal boron nitride crystals.  

\bibliographystyle{unsrt}
\bibliography{references}

% Supplementary material
\newpage
\clearpage
\renewcommand\thefigure{M\arabic{figure}}
\setcounter{figure}{0}

% Fig. M1
\begin{figure*}
    \centering
    \includegraphics{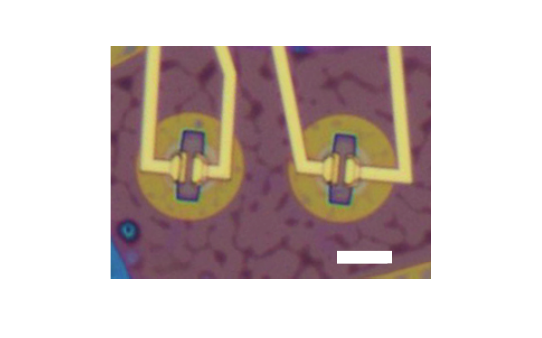}
    \caption{\textbf{Optical micrograph of the device.} Scale bar corresponds to 5 $\mu$m.}
    \label{fig:supp:optical_image}
\end{figure*}

% Fig. M2
\begin{figure*}
    \centering
    \includegraphics{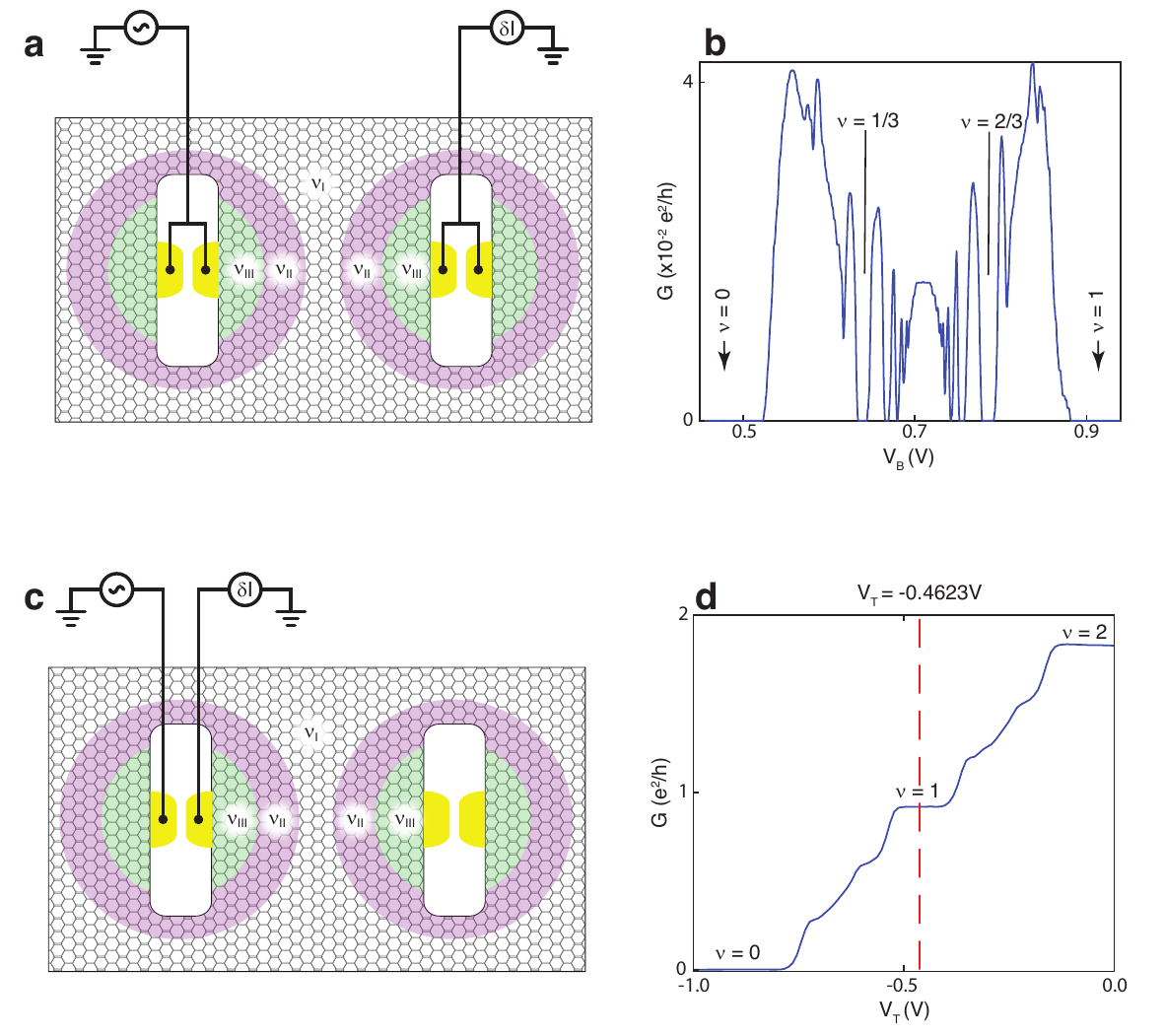}
    \caption{\textbf{Methods of local transport measurements.} 
    % subfigure a
    \textbf{a}, Inter-island transport measurement, referred to as $G$ in the main text. Because there is no edge state that directly couples charge across the pink region, non-zero $G$ is measured only when $\nu_\text{II}$ is tuned to a bulk conducting filling factor. Moreover, $|\nu_\text{II} - \nu_\text{I}|$ should be less than 1 to avoid fringe field-induced insulating phase between the pink and white region.
    % subfigure b
    \textbf{b}, Inter-island transport trace at $B_\perp = 8$T between $\nu=0$ to $\nu=1$.
    % subfigure c
    \textbf{c}, Intra-island transport. The two contacts on the right-hand side are kept open, and the two terminal resistance within a single island is measured. 
    % subfigure d
    \textbf{d}, Typical intra-island transport trace at $B_\perp = 8$T. The silicon gate voltage is fixed at 10V. The result shows that by fixing the silicon gate at 10V and the top gate at -0.4623V, we can set $\nu_\text{III} \geq 2$ and $\nu_\text{II} = 1$.
   } 
    \label{fig:supp:wiring}
\end{figure*}

%%%%%%%%%%%%%%%%%%%%%%%%%%%%%

\renewcommand\thefigure{S\arabic{figure}}
\setcounter{figure}{0}

% Fig. S1
\begin{figure*}
    \centering
    \includegraphics{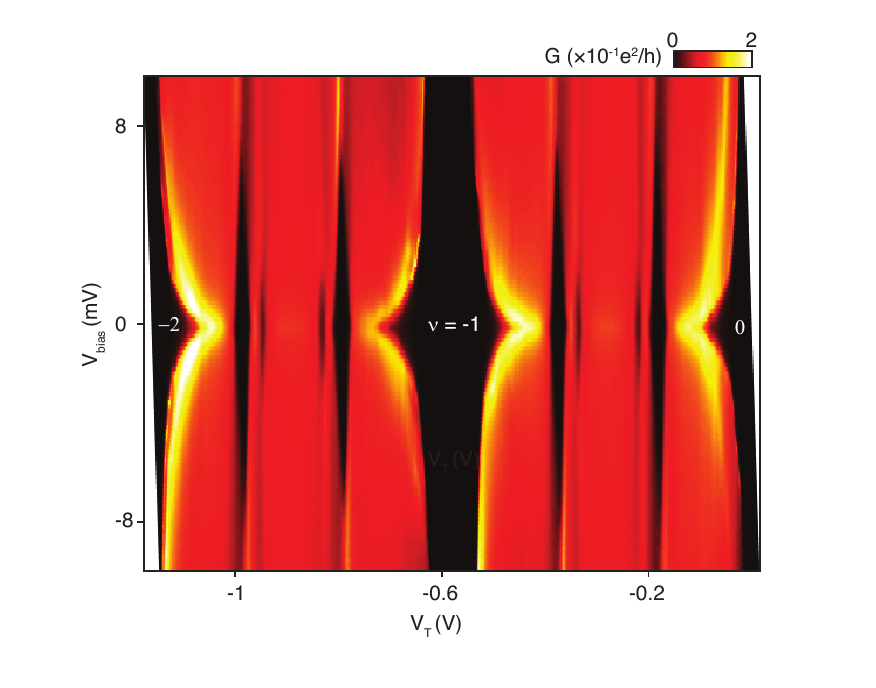}
    \caption{\textbf{Non-linear transport near $\nu = -1$.}
    Differential conductance $G$ as a function of gate voltage $V_\text{T}$ and source-drain bias voltage $V_\text{bias}$ measured in a second device at $B_\perp = 8$T. Strong nonlinearity is observed near all integer fillings.  
    }
    \label{fig:supp:nonlinear}
\end{figure*}

% Fig. S2
\begin{figure*}
    \centering
    \includegraphics{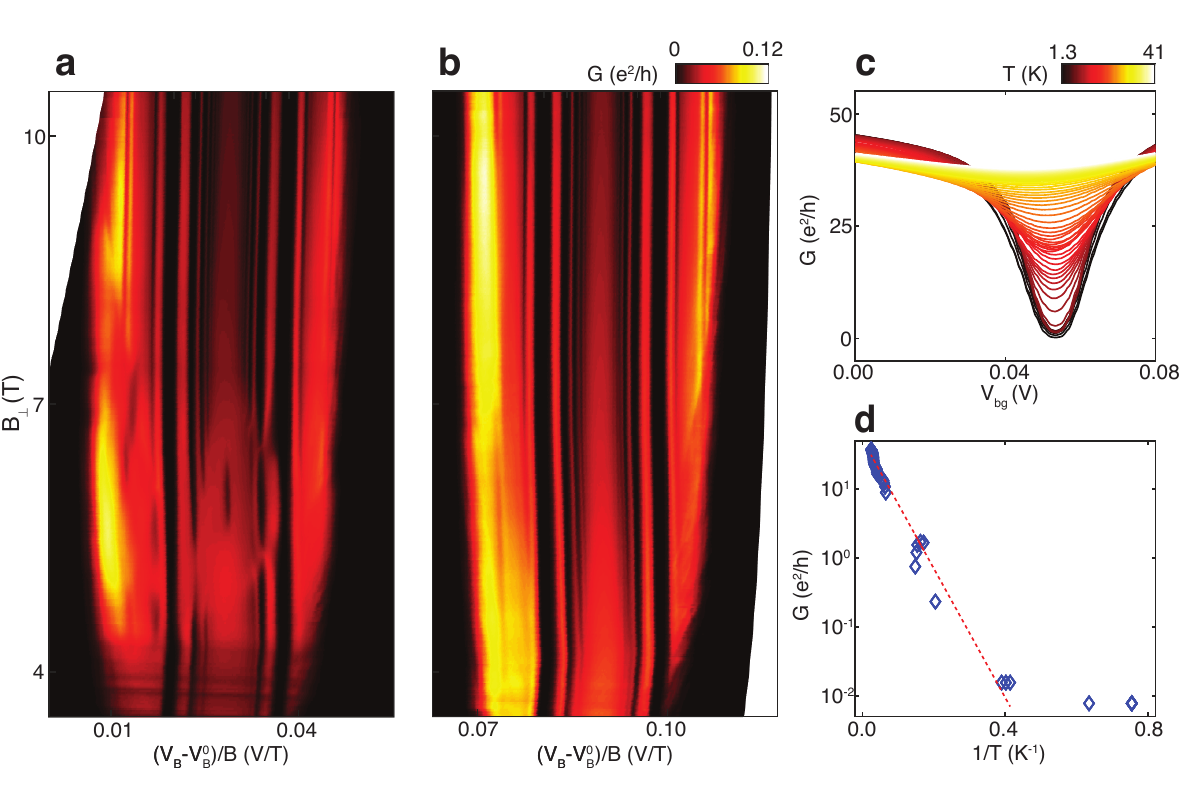}
    \caption{\textbf{Transport characterization of the device.}
    % subfigure a and b
    \textbf{a} Conductance as function of carrier density and magnetic field fo $0<\nu<1$ and \textbf{b} for $0<\nu<2$.  The device shows a series of FQH phase transitions for $0<\nu<1$, including the appearance of an even denominator state at $\nu=1/2$ over a narrow range of magnetic fields, all of which are absent for $1<\nu<2$\cite{zibrov_even-denominator_2018}.  
    % subfigure c
    \textbf{c}, Conductance near the charge neutrality point at $B=0$ at different temperatures. 
    % subfigure d
    \textbf{d}, Arrhenius plot of the conductance at the charge neutrality point, showing simply activated behavior consistent with a substrate induced sublattice splitting\cite{amet_insulating_2013,hunt_massive_2013}. The thermal activation gap is measured to be 3.7 meV.}
    \label{fig:supp:transport}
\end{figure*}

% Fig. S3
\begin{figure*}
    \centering
    \includegraphics{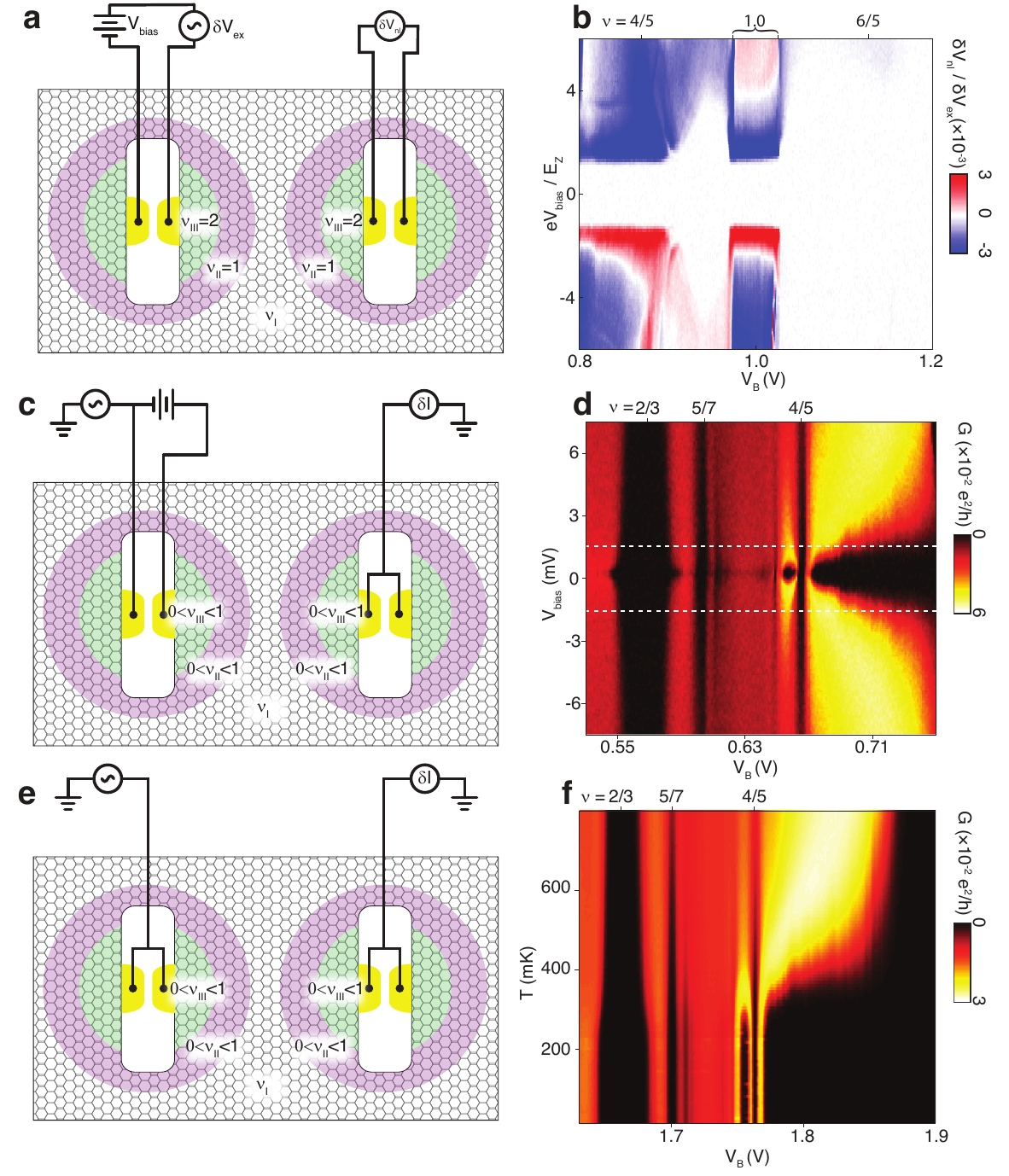}
    \caption{\textbf{Joule heating and its effect on the nonlocal mesurement.} 
    % subfigure a
    \textbf{a}, Schematic for measurements of the nonlocal response.  A DC bias is applied between the two left contacts, which are tuned to have an intra-island conductance $G=e^2/h$.  This results in a power dissipated via Joule heating of $P=V_\text{bias}^{2}G$. At the Zeeman threshold, $V_{bias}=E_Z/e$, resulting in $P=E_Z^2/h\approx .52 \text{pW}\times B_\text{[Tesla]}^2$.     % subfigure b
    \textbf{b},  As described in the main text, strong magnon transport signals are observed only for $V_{bias}>E_Z/e$, where the Joule heating power, dissipated directly on the device, leads to nonnegligible heating of the electron system. Data shown here are measured at B=10T. 
    % subfigure c
    \textbf{c}, Experimental setup for determining the effects of Joule heating.  In this setup, a bias voltage is applied across one island, as in the nonlocal measurement, but region II is tuned to be conducting.  This allows current to flow to the other island and the local conductivity to be measured as a function of both $V_B$ and the bias voltages.  Note that the small \textit{local} conductance across region I, $G\approx .01 e^2/h$, ensures that only minimal power is evolved from DC current flow across the sample bulk, providing a good proxy for heating in the nonlocal configuration. 
    % subfigure d
    \textbf{d}, Local conductivity measured in the setup described in panel c as a function of $V_{bias}$ and $V_B$ at B=13.5T.  As is evident in the data, applied bias leads to rapid degradation of the insulating phases associated with electron solids near $\nu_I=2/3$, $4/5$, and $1$.  At the $V_{bias}= \pm E_Z/e$ (indicated by dashed white lines), only a portion of the electron solid proximal to $\nu=1$ still shows insulating behavior.  
    % subfigure e
    \textbf{e}, Experimental setup for local conductance measurement. 
    % subfigure f
    \textbf{f}, Local conductance as a function of temperature at B=13.5T.  Comparing with the bias dependence shown in panel d, we conclude that bias-induced heating bring the electron temperature to $T\approx 400 mK$ at $eV_{bias}\approx E_Z$.   Notably, because the Zeeman energy sets the minimal bias voltage required to launch magnons, the effects of Joule heating grow significantly worse at high fields, scaling as $B^2$.  In contrast, the Coulomb interaction scales as $e^2/l_b \propto B_{\perp}^{1/2}$. The competition between the Coulomb interaction and sample heating limits the range of $B_{\perp}$ in which the electron solid phase exists when $eV_\text{bias}>E_\text{Z}$. Specifically, at large $B_{\perp}$, the  electron solid melts when $eV_\text{bias}>E_\text{Z}$ due to heating. We believe bias-induced Joule heating is primarily responsible for our observation of possible fractional skyrmion solids only at  $B_{\perp} = 7T$, but not at larger $B_{\perp}$ where they are more developed.  Interestingly, the low-B limit in our experiment is set not by the absence of correlation physics in region I, but rather by the breakdown of quantum Hall ferromagnetism in the much more disordered region II, which is exposed through the hBN dielectric to a disordered hBN/$\text{SiO}_\text{2}$ interface.  It may thus be possible to significantly improve on the current measurements by constructing devices in which all three necessary regions are of comparable sample quality.  
    }
    \label{fig:supp:heating}
\end{figure*}

% Fig. S4
\begin{figure*}
    \centering
    \includegraphics{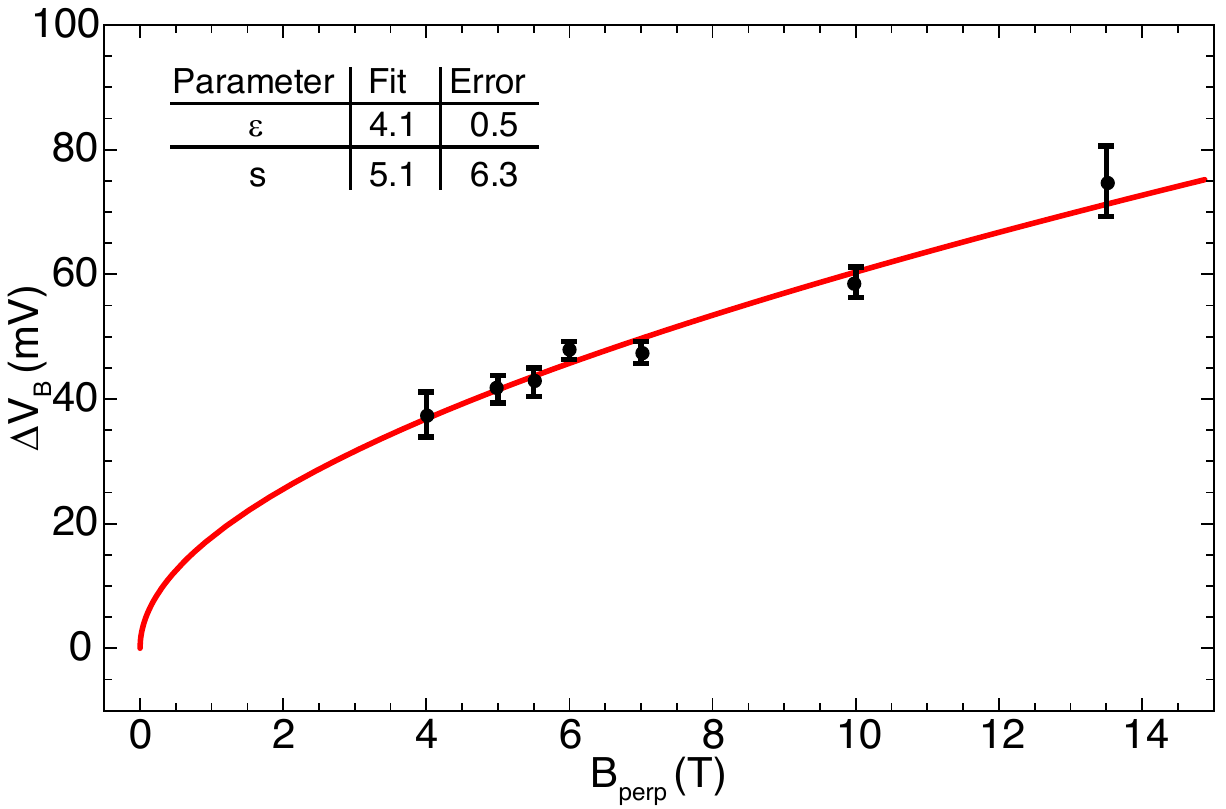}
    \caption{\textbf{Comparison of measured $\nu_\text{I}=1$ gap to theoretical calculation.}
Symbols indicate the measured width, $\Delta V_\text{B}$, corresponding to large nonlocal response, plotted for a variety of magnetic fields.  Theoretically, $\Delta V_\text{B}$ should directly give the energy gap of the quantum Hall ferromagnet ($\Delta$) via the relation $e \Delta V_\text{B}=\Delta$. $\Delta$ itself is predicted to follow the relation $\Delta=\sqrt{\frac{\pi}{2}}\frac{e^2}{\epsilon \ell_B} +s g\mu_B B_T$ (here $\epsilon$ is the dielectric constant and $s=2K+1$ tracks the number of excess spin flips, $K$, within a skyrmion excitation). The red curve represents a best fit to the measured data with $\epsilon$ and $s$ as free parameters. Due to the small contribution of the Zeeman energy to the total gap, $s$ is not well constrained, with a best fit of $5.1\pm6$. This fit is thus unable to discriminate between small skyrmions and single spin flip excitations, but is consistent with theoretical predictions of small skyrmions in the experimental range of $E_Z/E_C$.  The best fit parameter $\epsilon=4.1\pm.5$ agrees, to within error, with the expected value of $\epsilon=\sqrt{3\times 6.6}\approx 4.4$, consisting of the geometric average of the in- and out-of-plane dielectric constants of the encapsulating hBN layers. Note that Landau level broadening was recently measured to be $<1$ meV in similar devices\cite{polshyn_quantitative_2018,zeng_high_2018}, and is thus negligible on the scale of the $\nu=1$ energy gap over the measured range of magnetic fields.
}
\label{fig:supp:nu1gap}
\end{figure*}

\end{document}